\documentclass[twocolumn,prd,longbibliography]{revtex4}
\usepackage{graphicx}

\newcommand{\R}{{\bf R}}
\newcommand{\Z}{{\bf Z}}

\newcommand{\CI}{{\cal I}}

\newcommand{\CO}{{\cal O}}
\newcommand{\CT}{{\cal T}}

\newcommand{\CW}{{\cal W}}

\newcommand{\bk}{{\bf k}}
\newcommand{\bp}{{\bf p}}
\newcommand{\bq}{{\bf q}}
\newcommand{\bx}{{\bf x}}

\newcommand{\p}{\partial}

\newcommand{\be}{\begin{equation}}
\newcommand{\ee}{\end{equation}}
\newcommand{\bea}{\begin{eqnarray}}
\newcommand{\eea}{\end{eqnarray}}
\usepackage{graphicx}

\newcommand{\muir}{{\mu^{\ }_{\rm IR}}}
\begin{document}
\title{Cascading Multicriticality in Nonrelativistic Spontaneous Symmetry 
Breaking}
\author{Tom Griffin${}^a$, Kevin T. Grosvenor${}^{b,c}$, 
Petr Ho\v{r}ava${}^{b,c}$ and Ziqi Yan${}^{b,c}$}
\affiliation{\smallskip
${}^a$Blackett Laboratory, Department of Physics, Imperial College\\ 
London, SW7 2AZ, United Kingdom\smallskip\\
${}^b$Berkeley Center for Theoretical Physics and Department of Physics\\ 
University of California, Berkeley, California 94720-7300\smallskip\\
${}^c$Physics Division, Lawrence Berkeley National Laboratory\\ 
Berkeley, California 94720-8162}
\begin{abstract}\noindent Without Lorentz invariance, spontaneous global 
symmetry breaking can lead to multicritical Nambu-Goldstone modes with 
a higher-order low-energy dispersion $\omega\sim k^n$ ($n=2,3,\ldots$), 
whose naturalness is protected by polynomial shift symmetries.  Here we 
investigate the role of infrared divergences and the nonrelativistic 
generalization of the Coleman-Hohenberg-Mermin-Wagner (CHMW) theorem.  
We find novel cascading phenomena with large hierarchies between the 
scales at which the value of $n$ changes, leading to an evasion of the 
``no-go'' consequences of the relativistic CHMW theorem.  
\end{abstract}
\maketitle
\section{Introduction}

Some of the most pressing questions about the fundamental laws of 
the Universe (such as the cosmological constant problem, or the hierarchy 
between the Higgs mass and the Planck scale) can be viewed as puzzles of 
technical naturalness \cite{th}.  In this Letter, we study the interplay of 
technical naturalness with spontaneous symmetry breaking (SSB) in 
nonrelativistic systems.

SSB is ubiquitous in Nature.  For relativistic systems and global continuous 
internal symmetries, the universal features of SSB are controlled by the 
Goldstone theorem.  Much progress in SSB has also been achieved in the 
nonrelativistic 
cases, where the reduced spacetime symmetries allow a much richer behavior, 
still very much the subject of active research (see e.g.\ 
\cite{hbn,cb,n3,tb,hahi,msb,pol} and references therein).  
Important novelties emerge already in the simplest case of theories in the flat 
nonrelativistic spacetime $\R^{D+1}$ (covered with Cartesian coordinates 
$(t,\bx)$, 
$\bx\equiv(x^i,i=1,\ldots,D)$) and with the Lifshitz symmetries of spatial 
rotations and spacetime translations.  In such theories, the Nambu-Goldstone 
(NG) modes can be of two distinct types:  Type A, effectively described by a 
single real scalar $\phi(t,\bx)$ with a kinetic term quadratic in the time 
derivatives; or Type B, described by {\it two\/} scalar fields 
$\phi_{1,2}(t,\bx)$ which have a first-order kinetic term and thus form 
a canonical pair.  

In \cite{msb,pol}, we showed that this Type A-B dichotomy is further refined  
into two discrete families, labeled by a positive integer $n$:  
Type A${}_n$ NG modes are described by a single scalar with dispersion 
$\omega\sim k^n$ (and dynamical critical exponent $z=n$), while Type B${}_{2n}$ 
modes are described by a canonical pair and exhibit the dispersion relation 
$\omega\sim k^{2n}$ (and dynamical exponent $z=2n$).  These  
two families are technically natural, and therefore stable under 
renormalization in the presence of interactions \cite{msb}.  As usual, such 
naturalness is explained by a new symmetry.  For $n=1$, the NG modes are 
protected by the well-known constant shift symmetry $\delta \phi(t,\bx)=b$.  
The $n>1$ theories enjoy shift symmetries by a degree-$P$ 
polynomial in the spatial coordinates \cite{msb}, 
\be
\delta\phi(t,\bx)=b+b_ix^i+\cdots+b_{i_1\ldots i_P}x^{i_1}\cdots x^{i_P},
\ee
with suitable $P$.  Away from the Type 
A${}_n$ and B${}_{2n}$ Gaussian fixed points, the polynomial shift symmetry is 
generally broken by most interactions.  The lowest, least irrelevant 
interaction terms invariant under the polynomial shift were systematically 
discussed in \cite{pol} (see also \cite{hj}).  Such terms are often highly 
irrelevant compared to all the other possible interactions that break the 
symmetry.  

Having established the existence of the multicritical Type A and B families 
of NG fixed points, in this Letter we study the dynamics of flows between such 
fixed points in interacting theories.  We uncover a host of novel phenomena, 
involving large, technically natural hierarchies of scales, protected again by 
the polynomial shift symmetries.  As a given theory flows between the 
short-distance and the long-distance regime, it can experience a natural 
cascade of hierarchies, sampling various values of the dynamical critical 
exponent $z$ in the process.  Such cascades represent an intriguing mechanism 
for evading some of the consequences of the relativistic 
Coleman-Hohenberg-Mermin-Wagner (CHMW) Theorem.  

\section{The Relativistic CHMW Theorem}

Recall that in relativistic systems, all NG bosons are of Type~A${}_1$, 
assuming that they exist as well-defined quantum objects.  Whether or not 
they exist, and whether or not the corresponding symmetry can be 
spontaneously broken, depends on the spacetime dimension.   This phenomenon 
is controlled by a celebrated theorem, discovered independently in condensed 
matter by Mermin and Wagner \cite{mermw} and by Hohenberg \cite{hoh}, and in 
high-energy physics by Coleman \cite{cole}.  We therefore refer to this 
theorem, in the alphabetical order, as the CHMW theorem.  

The relativistic CHMW theorem states that the spontaneous breaking of global 
continuous internal symmetries is not possible in $1+1$ spacetime dimensions.  
The proof is beautifully simple.  $1+1$ is the ``lower critical 
dimension,'' where $\phi$ is formally dimensionless at the Gaussian fixed 
point.  Quantum mechanically, this means that its propagator is 
logarithmically divergent, and we must regulate it by an infrared (IR) 
regulator $\muir$:
\bea
\label{iras}
\langle\phi(x)\phi(0)\rangle&=&\int\frac{d^2k}{(2\pi)^2}\frac{e^{ik\cdot x}}{k^2
+\mu_{\rm IR}^2}\\
\approx&-&\frac{1}{2\pi}\log(\muir|x|)+{\rm const.}+\CO(\muir|x|).\nonumber
\eea
The asymptotic expansion in (\ref{iras}), valid for $\muir|x|\ll 1$, 
shows that as we take $\muir\to 0$, the propagator stays 
sensitive at long length scales to the IR regulator.  We can still construct 
various composite operators from the derivatives and exponentials of $\phi$, 
with consistent and finite renormalized correlation functions in the 
$\muir\to 0$ limit, but the field $\phi$ itself does not exist as a quantum 
object.  Since the candidate NG mode $\phi$ does not exist, the 
corresponding symmetry could never have been broken in the first place,
which concludes the proof.  

\section{Nonrelativistic CHMW Theorem}

For nonrelativistic systems with Type A${}_n$ NG modes, we find 
an intriguing nonrelativistic analog of the CHMW theorem.  The scaling 
dimension of $\phi(t,\bx)$ at the A${}_n$ Gaussian fixed point, measured in 
the units of spatial momentum, is
\be
[\phi(t,\bx)]^{\ }_{{\rm A}{}_n}=(D-n)/2.
\ee
The Type A${}_n$ field $\phi$ is at its lower critical dimension when 
$D=n$.  Its propagator then requires an IR regulator.  There are many 
ways how to introduce $\muir$; for example, by modifying the dispersion 
relation of $\phi$, as in 
\be
\langle\phi(t,\bx)\phi(0)\rangle=\int\frac{d\omega\,d^D\bk}{(2\pi)^{D+1}}
\frac{e^{i\bk\cdot\bx-i\omega t}}{\displaystyle{\omega^2+|\bk|^{2D}
+\mu_{\rm IR}^{2D}}},
\ee
or 
\be
\langle\phi(t,\bx)\phi(0)\rangle=\int\frac{d\omega\,d^D\bk}{(2\pi)^{D+1}}
\frac{e^{i\bk\cdot\bx-i\omega t}}{\displaystyle{\omega^2+(|\bk|^2+\mu_{\rm IR}^2)^D}}.
\ee

Regardless of how $\muir$ is implemented, as we take $\muir\to 0$, 
the propagator again behaves logarithmically, both in space
\be
\langle\phi(t,\bx)\phi(0)\rangle\approx-\frac{1}{(4\pi)^{D/2}\Gamma(D/2)}
\log(\muir|\bx|)+\ldots
\ee
for $|\bx|^D\gg t$, and in time,
\be
\langle\phi(t,\bx)\phi(0)\rangle\approx-\frac{1}{(4\pi)^{D/2}D\Gamma(D/2)}
\log(\mu_{\rm IR}^Dt)+\ldots
\ee
for $|\bx|^D\ll t$.  The propagator remains sensitive to the IR regulator 
$\muir$.  Consequently, we obtain the {\it nonrelativistic 
multicritical CHMW theorem for Type A modes\/}:

{\it The propagator of the Type A${}_n$ would-be NG mode $\phi(t,\bx)$ at its 
lower critical dimension $D=n$ is logarithmically sensitive to $\muir$, and 
therefore $\phi(t,\bx)$ does not exist as a quantum mechanical object.  
Consequently, no spontaneous symmetry breaking with Type A${}_n$ NG modes is 
possible in $D=n$ spatial dimensions.}

By extension, this invalidates all Type A${}_n$ would-be NG modes with 
$n>D$, whose propagator would also be pathological at long distances.   

In contrast, the scaling dimension of the Type B${}_{2n}$ fields is 
\footnote{Here we have assumed that both components of the canonical Type B pair $\phi_{1,2}$ carry the same dimension.  The more general case would only require the sum of their dimensions to equal $D$, still preventing a logarithmic IR divergence in $\langle\phi_1\phi_2\rangle$.}
\be
[\phi_{1,2}(t,\bx)]^{\ }_{{\rm B}{}_{2n}}=D/2,
\ee
and the lower critical dimension is $D=0$.  Hence, in all spatial 
dimensions $D>0$, the Type B${}_{2n}$ NG modes are free of IR 
divergences and well-defined quantum mechanically for all $n=1,2,\ldots$.  
The {\it nonrelativistic multicritical CHMW theorem for Type B modes\/} then 
simply states that 
{\it the Type B${}_{2n}$ symmetry breaking is possible in any $D>0$ and for 
any $n=1,2,\ldots$.}

\section{Cascading Multicriticality}

Whereas in the relativistic case, all NG modes must always be of Type A${}_1$, 
in nonrelativistic systems the existence of the Type A${}_n$ and B${}_{2n}$ 
families allows a much richer dynamical behavior. 

For example, with the changing momentum or energy scales, a given NG mode 
can change from Type A${}_n$ (or B${}_{2n}$) to Type A${}_{n'}$ 
(or B${}_{2n'}$) with $n\neq n'$, or it could change from Type A to 
Type B.  The hierarchies of scales that open 
up in this process are naturally protected by the corresponding polynomial 
symmetries.  One of the simplest cases is the Type A${}_n$ scalar with $n>1$, 
whose polynomial shift symmetry of degree $P$ is broken at some momentum 
scale $\mu$ to the polynomial shift symmetry of degree $P-2$, by some small 
amount $\varepsilon\ll 1$.  This breaking modifies the dispersion relation to 
$\omega^2\approx k^{2n}+\zeta_{n-1}^2k^{2n-2}$, with 
$\zeta_{n-1}^2\approx\varepsilon\mu^2$.  Here, as in \cite{th}, we identify 
$\mu$ as the scale of naturalness.  At a hierarchically much smaller scale, 
$\mu_\times\equiv\mu\sqrt{\varepsilon}$, the system exhibits a crossover, 
from Type A${}_n$ above $\mu_\times$ to Type A${}_{n-1}$ below $\mu_\times$.  
The technical naturalness of the large hierarchy $\mu_\times\ll\mu$ is 
protected by the restoration of the polynomial shift symmetry of degree $P$ as 
$\varepsilon\to 0$.  

In the special case of $n=D$, this crossover from Type A${}_D$ to Type 
A${}_{D-1}$ yields an intriguing mechanism for evading the naive conclusion of 
our CHMW theorem.  For a large range of scales close to $\mu$, the would-be 
NG mode can exhibit a logarithmic propagator.  The hierarchically smaller 
scale $\mu_\times\ll\mu$ then serves as a natural IR 
regulator, allowing the NG mode to cross over to Type A${}_{D-1}$ at very long 
distances.  Therefore, the mode is well-defined as a quantum mechanical 
object, despite the large hierarchy across which it behaves effectively 
logarithmically.

An interesting refinement of this scenario comes from breaking the polynomial 
symmetries hierarchically, in a sequence of partial breakings, from a higher 
polynomial symmetry of degree $P$ to symmetries with degrees $P'<P$, $P''<P'$, 
$\ldots$, all the way to constant shift.  This gives rise to a cascading 
phenomenon, with a hierarchy of crossover scales $\mu\gg\mu'\gg\mu''\gg
\ldots$, separating plateaux governed by the fixed points with the dynamical 
exponent taking the corresponding different integer values.  Again, such 
cascading hierarchies are technically natural, and protected by the underlying 
breaking pattern of the polynomial symmetries.  

Before we illustrate this behavior in a series of examples, it is worth 
pointing out one very simple yet important feature of large hierarchies in 
nonrelativistic theories.  Consider a theory dominated over some range of 
scales by the dispersion relation $\omega\approx k^n$, with $n>1$.  If we open 
up a large hierarchy of momentum scales $\mu\gg\mu'$ (say by $N$ orders of 
magnitude), this hierarchy of momentum scales gets magnified into an even 
larger hierarchy (by $nN$ orders of magnitude) in energy scales.  

\section{Example 1: a Type-A Hierarchy}

The first model that we use to illustrate the hierarchy is a relatively 
well-known system in $2+1$ dimensions: the $z=2$ Gaussian model of a single 
Lifshitz scalar field $\phi(t,\bx)$, with a derivative 4-point 
self-interaction turned on \cite{gg,lubensky}:
\[
S_2=\frac{1}{2}\int dtd^2\bx\left\{\dot\phi^2-(\p^2\phi)^2
-c^2\p_i\phi\p_i\phi-g(\p_i\phi\p_i\phi)^2\right\}.
\]
This action contains all the marginal and relevant terms of the $z=2$ fixed 
point consistent with the constant shift symmetry and the reflection symmetry 
$\phi\to-\phi$.  At the $z=2$ Gaussian fixed point, $g$ is classically 
marginal, and breaks the polynomial shift symmetry of this fixed point to 
constant shift.  Quantum corrections at one loop turn $g$ marginally 
irrelevant \cite{gg}.  

This system allows a natural hierarchy of scales, stable under quantum 
corrections.  At the naturalness scale $\mu$, we can break the polynomial 
shift symmetry of the $z=2$ fixed point to constant shifts by a small amount 
$\varepsilon_0\ll 1$.  This implies $g\sim\varepsilon_0$ and 
$c^2\sim\varepsilon_0\mu^2$, relations which can be shown to be respected by 
the loop corrections.  In particular, $c^2$ can stay naturally small, much 
less than $\mu^2$.  The dispersion relation changes from $z=2$ close to the 
high scale $\mu$, to $z=1$ around the much lower scale 
$\mu_1\equiv\mu\sqrt{\varepsilon_0}\ll\mu$.  

\section{Example 2: a Type-A Cascade}

Our next example is a new model, which not only illustrates the cascading 
hierarchy with multiple crossovers, but also exhibits additional intriguing 
renormalization properties of independent interest.  

\begin{figure}
\includegraphics[angle=0,width=2.2in]{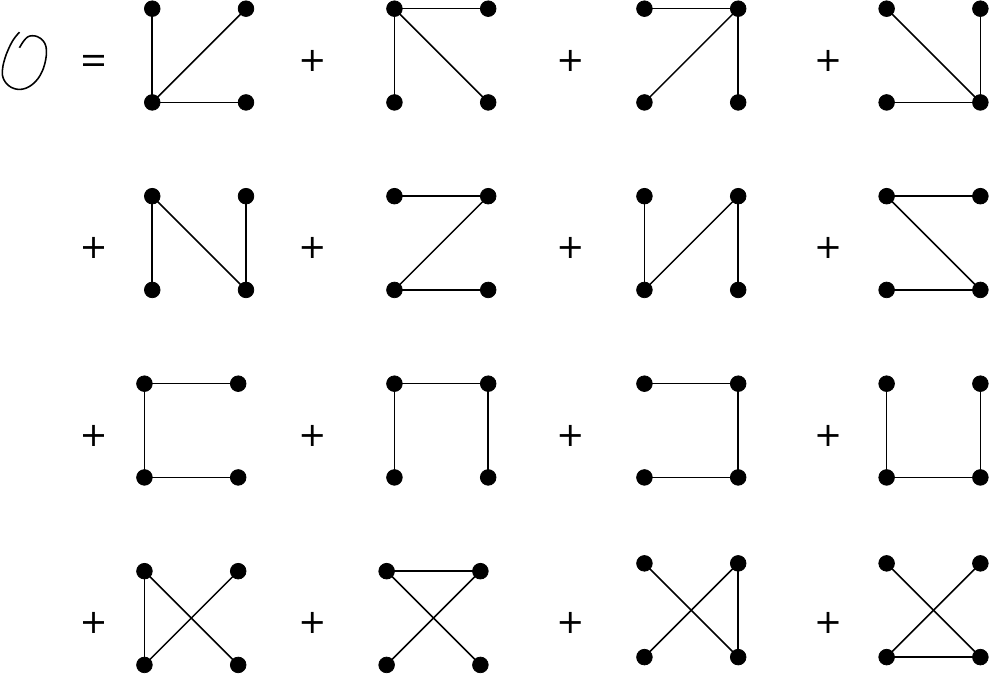}
\caption{\label{4by4}Graphical representation of the unique four-point invariant $\CO$ of the linear shift symmetry, as an equal-weight sum of all trees with distinguishable vertices (see \cite{pol}).  Each dot represents one copy of $\Phi$. Each link represents a contracted pair of derivatives $\p_i\ldots\p_i$ acting at the ends of the link.}
\end{figure}

We start with the Gaussian $z=3$ fixed point of a single scalar $\Phi(t,\bx)$ 
in $3+1$ dimensions, and turn on derivative self-interactions and relevant 
terms as in our previous example.  The free theory is
\bea
S_3&=&\frac{1}{2}\int dtd^3\bx\left\{\dot\Phi^2-
\zeta_3^2(\p_i\p_j\p_k\Phi)(\p_i\p_j\p_k\Phi)\right.\nonumber\\
&&\left.{}-\zeta_2^2(\p^2\Phi)^2-c^2\p_i\Phi\p_i\Phi\right\}.
\eea
At the classical level we can set $\zeta_3^2=1$ by the rescaling of space.  
The terms on the second line represent relevant Gaussian deformations away from 
the $z=3$ fixed point.  The spectrum of available self-interaction operators 
that are classically marginal or relevant at the $z=3$ Gaussian fixed point 
is much richer than in our $2+1$ dimensional example.  We shall again restrict 
our attention only to the operators even under $\Phi\to-\Phi$, and invariant 
at least under the constant shift.  Up to total derivatives, which we ignore, 
there are three independent marginal 4-point operators $\CO_4^{(a)}$, $a=1,2,3$, 
one marginal 6-point operator $\CO_6=(\p_i\Phi\p_i\Phi)^3$, and one relevant 
4-point operator 
\be
\CW=(\p_i\Phi\p_i\Phi)^2.
\ee
Among them, there is one unique operator $\CO$ invariant under the linear shift 
symmetry up to a total derivative:
\bea
\CO&=&4\p_i\p_j\p_k\Phi\p_i\Phi\p_j\Phi\p_k\Phi
+12\p_i\Phi\p_i\p_j\Phi\p_j\p_k\Phi\p_k\Phi\nonumber\\
&&{}=\ 4\ \vcenter{\hbox{\includegraphics[angle=0,width=.25in]{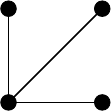}}}
+12\ \vcenter{\hbox{\includegraphics[angle=0,width=.25in]{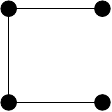}}}\ ,
\eea
cf.\ Fig.(\ref{4by4}).  This operator is classically marginal.

To construct our model, we start with the free theory $S_3$ and turn on just 
the unique linear-shift invariant self-interaction $\CO$, with coupling 
$\lambda$.  The Feynman rules of this model involve one 4-vertex, which 
can be simplified using the momentum conservation $\bk_4=-\sum_{I=1,2,3}\bk_I$  
to 
\bea
&&\ \!\!\!\!\!\!\vcenter{\hbox{\includegraphics[angle=0,width=.65in]{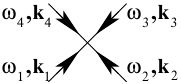}}}
=-i\lambda\left[k_1^2k_2^2k_3^2
+2(\bk_1\cdot\bk_2)(\bk_2\cdot\bk_3)(\bk_3\cdot\bk_1)\right.\nonumber\\
&&{}\quad\left.{}-k_1^2(\bk_2\cdot\bk_3)^2-k_2^2(\bk_3\cdot\bk_1)^2
-k_3^2(\bk_1\cdot\bk_2)^2\right].
\eea
Note that in this vertex, each momentum appears quadratically, with no 
subleading terms.  We can write it even more compactly with the use 
of the fully antisymmetric $\epsilon_{ijk}$ tensor: if for any three momenta 
$\bk,\bp,\bq$ we define $[\bk\bp\bq]\equiv\epsilon_{ijk}k_ip_jq_k$, our 
vertex becomes simply 
\be
\label{vertx}
-i\lambda[\bk_1\bk_2\bk_3]^2.
\ee
This simple vertex structure is intimately related to the underlying 
symmetries:  When translated into momentum space, the linear shift symmetry 
$\delta\phi(t,\bx)=b_ix^i+b$ becomes a shift of the Fourier modes 
$\phi(t,\bk)$ by $b_i(\p/\p\bk_i)\delta(\bk)+b\delta(\bk)$.  Acting with this 
symmetry on any of the legs of the vertex produces zero, as the vertex is 
purely quadratic in each of the outside momenta. 

{\bf Quantum properties.}  This model has intriguing renormalization group 
properties, which will be discussed in detail elsewhere \cite{nrr}.  First, 
note that $\lambda$ satisfies a non-renormalization theorem:  Consider the 
$2N$-point function of $\Phi$, with $N>1$, and with external momenta $\bk_1, 
\ldots \bk_{2N}$.  Any 1PI diagram will be of the form $k^{i_1}_1k^{j_1}_1\cdots
k^{i_{2N}}_{2N}k^{j_{2N}}_{2N}\cdot\CI_{i_1j_1\ldots i_{2N}j_{2N}}(\bk_1,\ldots,\bk_{2N})$.  
The factor 
$\CI$ has no ultraviolet divergences, and with $c^2\neq 0$ or $\zeta_2^2\neq 0$ 
it approaches a finite value at $\bk_1=\ldots=\bk_{2N}=0$.  The special case of 
$N=2$ implies that $\lambda$ does not get renormalized, at any loop order.  
Note also that none of the operators $\CW$, $\CO_6$ or $\CO_4^{(a)}$ that would 
break the linear shift symmetry are generated by the loop corrections.  

The remarkable non-renormalization of $\lambda$ does not imply that the 
effective self-interaction strength would be independent of scale:  There 
is a non-trivial renormalization of the 2-point function.  While the one-loop 
diagram 
$\vcenter{\hbox{\includegraphics[angle=0,width=.4in]{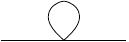}}}$ 
is identically zero, the two-loop 
diagram $\vcenter{\hbox{\includegraphics[angle=0,width=.4in]{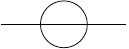}}}$
gives the generic behavior which persists at higher loops:  There is no 
wave-function renormalization, no renormalization of $c^2$, the loop 
corrections to $\zeta_2^2$ are quadratically divergent, and those to 
$\zeta_3^2$ diverge logarithmically \cite{nrr}.  This log divergence 
effectively corrects the dynamical exponent of the ultraviolet 
fixed point away from the classical value $z=3$.  The modified scaling in turn 
implies that the theory becomes effectively strongly coupled at some finite 
scale $\mu_s$.  

Having understood the quantum corrections, we can now study cascading 
hierarchies of symmetry breaking in this model, and confirm their technical 
naturalness.  At some high scale $\mu$, which will be our naturalness scale, 
and which we keep below $\mu_s$, consider the following hierarchical breaking 
of polynomial symmetries:  First, break the $P=4$ symmetry of the $z=3$ 
Gaussian fixed point to the $P=2$ symmetry of the $z=2$ fixed point by some 
small amount $\varepsilon_2\ll 1$.  Then break $P=2$ to $P=1$ by an even 
smaller amount $\varepsilon_1\ll\varepsilon_2$.  This pattern corresponds to 
\be
\label{hier}
\zeta_3^2\approx 1,\quad \zeta_2^2\approx\mu^2\varepsilon_2,\quad
c^2\approx\mu^4\varepsilon_1,\quad \lambda\approx\varepsilon_1.   
\ee
The dispersion relation cascades from $z=3$ at high energy scales, 
to $z=2$ at intermediate scales, to $z=1$ at low scales. 
\footnote{We can also break 
the linear shift symmetry to constant shifts, by some amount 
$\varepsilon_0\ll\varepsilon_1$.  This would generate the remaining classically 
marginal operators $\CO^{(a)}_4$, $\CO_6$ and the relevant operator $\CW$, with 
coefficients of order $\varepsilon_0$ in the units of $\mu$.}
Both the large hierarchies in (\ref{hier}) and the cascading behavior of the 
dispersion relation are respected by all loop corrections, and therefore are 
technically natural.   This follows by inspection from the properties of the 
quantum corrections discussed above.   

\section{Example 3:\break a Type-A/Type-B Hierarchy} 

So far, we focused on the cascading mechanism in the Type A case.  Type B 
systems can form their own hierarchies, in the obvious generalization of the 
Type A cascades exemplified above.  There is no analog of the lower critical 
dimension and the CHMW limit on $n$.  Type A NG modes can also exhibit a flow 
to Type B.  This behavior, albeit not new (see e.g.\ \cite{anton}), can be 
embedded as one step into the more general technically natural hierarchies of 
Type A or B as discussed above.  In particular, the crossover to Type B can 
provide a new IR regulator of the Type A cascade at the lower critical 
dimension.  

We shall illustrate this on the simplest Type A${}_1$ example, 
although the full story is, of course, more general.  
Consider two would-be Type A NG fields, $\phi_{1,2}(t,\bx)$, in the 
vicinity of the $z=1$ Gaussian fixed point
\[
S_1=\frac{1}{2}\int dtd^D\bx\left\{\dot\phi_1^2+\dot\phi_2^2-c_1^2(\p_i\phi_1)^2
-c_2^2(\p_i\phi_2)^2\right\}.
\]
For simplicity, we will set $c_1=c_2=1$, although this is not necessary 
for our argument.  Besides the rotations and translations of the two scalars, note two independent $\Z_2$ symmetries -- the field reflection $R$: $(\phi_1,\phi_2)\to(\phi_1,-\phi_2)$, and the time reversal $\CT$: $t\to-t$, $\phi_{1,2}(t,\bx)\to\phi_{1,2}(-t,\bx)$.  We can now turn on the Type B kinetic term, 
\be
S'=S_1+\Omega\int dtd^D\bx\left(\phi_1\dot\phi_2-\phi_2\dot\phi_1\right).
\ee
$\Omega$ now provides an IR regulator for the propagator.  At that scale, the field reversal $R$ and the time reversal $\CT$ are broken to their diagonal subgroup.  At energy scales below $\Omega$, one of the would-be Type A NG modes survives and turns into the Type B NG mode, while the other would-be Type A mode develops a gap set by $\Omega$.  Note that in $1+1$ dimensions, the ``no-go'' consequences of the relativistic CHMW theorem are again naturally evaded by this hierarchy: a NG mode exists quantum mechanically after all, and symmetry breaking is possible, despite the fact that above the scale $\Omega$, the two would-be Type A modes exhibit the logarithmic two-point function suggesting that symmetry breaking may not be possible.

The hierarchy between the Type A and Type B behavior is also protected by symmetries.  In fact, the system has multiple symmetries that can do this job.  One can rely on the breaking pattern of the discrete symmetries $R$ and $\CT$ mentioned above.  If the Type A system is Lorentz invariant, one can use Lorentz symmetry breaking to protect small $\Omega$.  More interestingly, without relying on the discrete or Lorentz symmetries, one can introduce a shift symmetry linear in time, $\delta\phi_{1,2}=b_{1,2}t$.  While the Type A kinetic term is invariant under this symmetry, the Type B kinetic term is not.  Breaking the linear shift symmetry to constant shifts allows the Type-A/Type-B crossover scale to be hierarchically smaller than the naturalness scale. 

\section{Conclusions and Outlook}

We have seen that the multicritical Type A${}_n$ and B${}_{2n}$ NG modes can experience technically natural cascading hierarchies of scales, protected by a hierarchy of polynomial shift symmetries.  Perhaps the most interesting case is Type A with $n=D$, which according to our CHMW theorem exhibits logarithmic sensitivity to the IR regulator.  In the relativistic case, this would prevent the symmetry breaking.  We have shown that the Type A${}_D$ modes can experience a cascade to Type A${}_n$ with $n<D$ (or to Type B), which provides a natural IR regulator, thus making the symmetry breaking possible after all.  

Our original motivation for this study of technical naturalness 
and hierarchies in SSB came from quantum gravity and high-energy 
physics \cite{msb,pol}, especially in the context of nonrelativistic gravity 
\cite{mqc,lif}.  Besides extending our understanding of the general 
``landscape of naturalness,'' we expect that our results could find their most 
immediate applications in two other areas:  In condensed 
matter physics, and in effective field theory of inflationary cosmology 
\cite{efti,eftiw,infst}.  
Both areas treat systems with nonrelativistic, Lifshitz-like symmetries  
similar to ours.  In condensed matter, the multicriticality of NG modes will 
affect their thermodynamic and transport properties; for example, the 
Type A${}_D$ modes at the lower critical dimension will exhibit specific heat 
linear in temperature $T$ over the range of $T$ dominated by the $z=D$ 
dispersion (up to $\log T$ corrections due to self-interactions).  In the 
context of inflation, our self-interacting scalar field theories represent a  
new nonrelativistic variation on the theme of the Galileon \cite{ggnrt}, 
an extension of the $z=2$ ghost condensate \cite{ghst1,ghst}, and of the $z=3$ 
cosmological scalar theory of Mukohyama \cite{z3s,shinjire}.

{\bf Acknowledgements:}  We wish to thank Christopher Mogni and 
Rikard von Unge for useful discussions.  This work has been supported by NSF 
Grant PHY-1214644 and by Berkeley Center for Theoretical Physics.  
\bibliography{cmu}
\end{document}